\documentclass[sigconf]{acmart}

\newif\ifworkinprogress

\usepackage[hide]{chato-notes}

\newif\ifworkinprogressomments
\ifworkinprogress
    \newcommand{\mst}[1]{\textbf{\color{olive}/* #1 (mst) */}}
    \newcommand{\lesp}[1]{\textbf{\color{blue}/* #1 (lesp) */}}
    \newcommand{\fle}[1]{\textbf{\color{orange}/* #1 (fle) */}}
    \newcommand{\mmu}[1]{\textbf{\color{green}/* #1 (mmu) */}}
    \newcommand{\swa}[1]{\textbf{\color{red}/* #1 (swa) */}}
\else
    \newcommand{\mst}[1]{}
    \newcommand{\lesp}[1]{}
    \newcommand{\fle}[1]{}
    \newcommand{\mmu}[1]{}
    \newcommand{\swa}[1]{}
\fi

\newcommand{\para}[1]{\smallskip\noindent\textbf{#1}}
\newcommand{\bioportal}{{BioPortal}}

\newcommand{\ontoportation}{{HopPortation}}   
\newcommand{\hoprank}{{Hop\-Rank}}  
\newcommand{\khop}{{k-hop}}  
\newcommand{\nontologies}{{11}}
\newcommand{\yearstudy}{{2015}}
\newcommand{\pagerank}{{PageRank}}
\newcommand{\markovchain}{{Markov chain}}
\newcommand{\hopportation}{{HopPortation}}

\newcommand{\gravi}{{Gravitational}}

\newcommand{\totaltransitions}{{$133K$}} 
\newcommand{\percentagetransitions}{{$80\%$}}

\newcommand{\loincN}{{$175K$}}
\newcommand{\loincE}{{$153K$}}
\newcommand{\loincCC}{{$74K$}}

\newcommand{\nnavitypes}{{$7$}}
\newcommand{\nomodels}{{$7$}}

\newcommand{\validtrans}{{$133K$}} 

\newcommand{\sessionbreak}{{$60$ minutes}}
\newcommand{\nsessions}{{$336K$}} 
\newcommand{\minsession}{{$2$}}
\newcommand{\mintrans}{{$1000$}}

\newcommand{\meddranodes}{{$23K$}} 
\newcommand{\meddratrans}{{$43K$}} 
\newcommand{\hoprankoutperfoms}{{89\%}}

\usepackage{booktabs} 
\usepackage[utf8]{inputenc}
\usepackage{url}
\usepackage{xcolor}
\usepackage{graphicx}
\usepackage{numprint}
\usepackage{graphicx}
\usepackage{amsmath}
\usepackage{multirow}
\usepackage[center]{subfigure}
\usepackage{cleveref}
\usepackage{enumitem}
\usepackage{booktabs}
\usepackage{balance}
\usepackage{titlecaps}
\usepackage[skip=1pt]{caption}

\copyrightyear{2019}
\acmYear{2019} 
\setcopyright{iw3c2w3}
\acmConference[WWW '19]{Proceedings of the 2019 World Wide Web Conference}{May 13--17, 2019}{San Francisco, CA, USA}
\acmBooktitle{Proceedings of the 2019 World Wide Web Conference (WWW '19), May 13--17, 2019, San Francisco, CA, USA}
\acmPrice{}
\acmDOI{10.1145/3308558.3313487}
\acmISBN{978-1-4503-6674-8/19/05}

\begin{document}

\title[\hoprank]{\hoprank: How Semantic Structure Influences Teleportation in \pagerank~(A Case Study on \bioportal)}

\author{Lisette Esp\'{i}n-Noboa}
\affiliation{\institution{GESIS \& Uni. Koblenz-Landau}}
\email{lisette.espin@gesis.org}

\author{Florian Lemmerich}
\affiliation{\institution{RWTH Aachen University}}
\email{florian.lemmerich@cssh.rwth-aachen.de}

\author{Simon Walk}
\affiliation{\institution{Detego GmbH}}
\email{s.walk@detego.com}

\author{Markus Strohmaier}
\affiliation{\institution{RWTH Aachen University \& GESIS}}
\email{markus.strohmaier@cssh.rwth-aachen.de}

\author{Mark Musen}
\affiliation{\institution{BMIR-Stanford}}
\email{musen@stanford.edu}

\renewcommand{\shortauthors}{Esp\'{i}n-Noboa et al.}

\begin{abstract}
This paper introduces \hoprank, an algorithm for modeling human navigation on semantic networks. 
\hoprank~leverages the assumption that users know or can see 
the whole structure of the network.
Therefore, besides following links, they also follow nodes at \emph{certain distances} (i.e., \khop~neighborhoods), and not at random as suggested by \pagerank, which assumes only links are known or visible. 
We observe such preference towards \khop~neighborhoods on \bioportal, one of the leading repositories of 
biomedical ontologies on the Web. In general, users navigate within the vicinity of a concept. But they also ``jump'' to distant concepts less frequently. We fit our model on \nontologies~ontologies
using the transition matrix of clickstreams, and show that semantic structure can influence teleportation in \pagerank.
This suggests that users---to some extent---utilize know\-ledge about the underlying structure of ontologies, and leverage it 
to reach certain pieces of information.
Our results help the development and improvement of user interfaces for ontology exploration.
\end{abstract}

\ccsdesc[500]{Information systems~Content ranking}
\ccsdesc[300]{Mathematics of computing~Exploratory data analysis}
\ccsdesc[100]{Information systems~Browsers}

\keywords{Biased random walker; \pagerank; \khop~neighborhood; \bioportal}

\maketitle

\begin{figure*}[t]
    \centering
    \subfigure[Transitions]{
        \label{fig:toy-example:sequence}
        \includegraphics[width=0.20\linewidth]{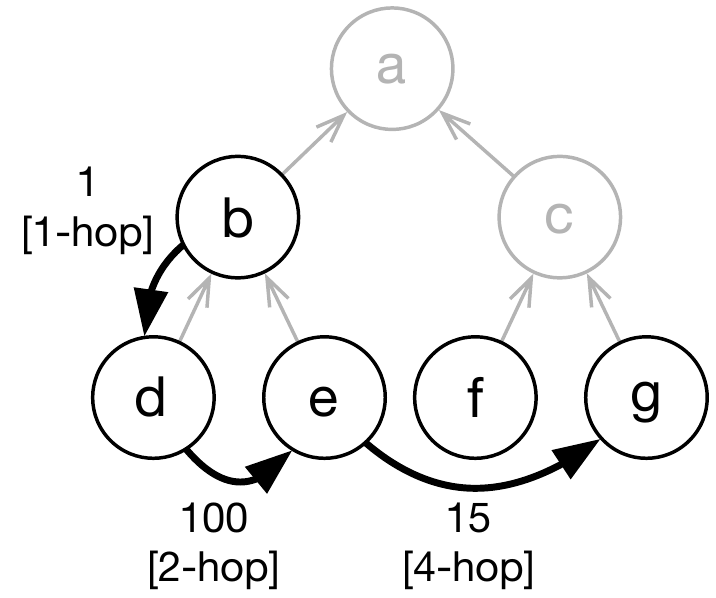}}
    \subfigure[HopPortation Vector]{
        \raisebox{7mm}{
        \label{fig:toy-example:hopportation}
        \includegraphics[width=0.30\linewidth]{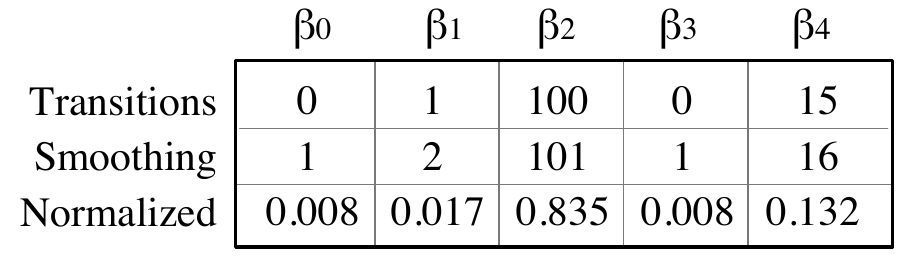}}}
    \subfigure[Transition Probabilities]{
        \label{fig:toy-example:probabilities}
        \includegraphics[width=0.35\linewidth]{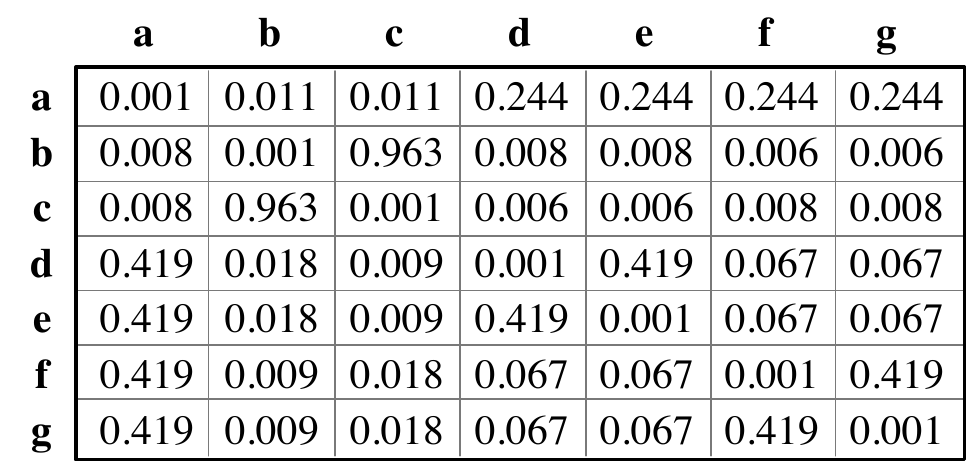}}
    \setlength{\belowcaptionskip}{-5pt} 
    \caption{\textbf{\hoprank~on semantic networks}. This example illustrates an instance of navigation on an ontology. (a) Shows the underlying network composed by seven concepts (a-g) and six \emph{isASubClassOf} relationships (straight-thin grey arrows). Transitions (curved-thick black arrows) are labeled by the actual number of transitions between concepts, as well as the [\khop] distance (i.e., shortest path) between them. (b) Illustrates how the \hopportation~vector $\vec{\beta}$ is built using transition counts per \khop. (c) Shows the transition probabilities inferred by \hoprank, see \Cref{eq:hoprank}.}
    \label{fig:toy-example}
\end{figure*}

\section{Introduction}

Ontology Engineering and Ontology Learning are two branches of the Semantic Web whose aim is to accurately build and curate ontologies.
The former studies new techniques to improve collaboration among humans while editing ontologies \cite{tudorache2008web,walk2015understanding}, and the latter introduces new methodologies and algorithms to automatically create ontologies by crawling the Web \cite{auer2007dbpedia,suchanek2007yago}. 
These efforts represent significant advances in the development of knowledge bases, which represent facts about the real world (e.g., people, diseases). 
However, there is little knowledge about 
how users consume such ontologies
on the Web. 
To this end, Walk et al. studied how users browse \bioportal~\cite{walk2017users}.
Their findings suggest that some ontologies influence the way users interact with the website.
However, how users navigate through the ontology structure (i.e., from one concept to another) remains unclear.

\para{Problem Statement:} 
In this paper, we study the influence of semantic structure on \emph{teleportation} (i.e.,  jumping to any node chosen at random) in PageRank. For example, consider the ontology shown in \Cref{fig:toy-example:sequence}, where nodes represent \emph{classes} (a.k.a. concepts) and edges \emph{isASubClassOf} relationships. 
On \bioportal, ontologies are shown vertically as hierarchical trees,
and concepts can be explored using the \emph{expand-on-demand} principle. This means that only top level concepts are shown first, and then users are able to expand and collapse as many concepts as they need at any level of the ontology. In other words, users can use and therefore are potentially aware of a \emph{virtually fully connected network} in all stages of navigation.
Previous studies \cite{page1999pagerank,xing2004weighted} have modeled user navigation using \pagerank. However, these assume that navigation paths are constrained by links and random teleportation. In our scenario, where the whole structure of an ontology can be visualized at any time, we believe that teleportation is not fully random, but rather biased towards \khop~neighborhoods.

\para{Approach:}
Motivated by previous studies on information foraging \cite{pirolli1999information,chi2000scent,card2001information,chi2001using}, decentralized search \cite{kleinberg2002small,helic2013models}, and \pagerank~\cite{page1999pagerank,haveliwala2003topic,xing2004weighted,gorman2004google,abbott2015random}, we propose \hoprank, a method for modeling transitions across \khop~neighborhoods on semantic networks. The key idea of this work relies on the \ontoportation~vector $\vec{\beta}$, which defines the probabilities of transitioning to each \khop~neighborhood. From the \pagerank~point of view, we can say that teleportation 
is not fully random, and the probability of following the structure of a page is not based only on one parameter (i.e., probability of following links), but on $k$ parameters, representing all \khop~neighborhoods \emph{reached from the current page}. Technically, we pass the \ontoportation~vector to a random walker to make biased decisions on which neighborhood to go next. Once this decision is made, the random walker uniformly chooses a concept within that neighborhood. 

\break
\para{Contributions:}
The contributions of this paper are: 
\begin{enumerate}
    \item We empirically show how users leverage the structure of the ontologies on \bioportal~by quantifying the proportion of transitions per \khop~neighborhood. 
    \item We propose \hoprank, an algorithm for modeling human navigation on semantic networks.
    \item We demonstrate that \hoprank~outperforms traditional navigation and popularity-based models on \bioportal, especially when users  browse ontologies directly without search.
    \item We make an implementation of this approach openly available on the Web \cite{hoprank}.
\end{enumerate}

\section{Related Work}
\label{sec:relatedwork}
\bioportal~provides users with a \emph{tree-like explorer} and a \emph{local search} engine to navigate ontologies.
In addition, concepts can be \emph{expanded on demand} to see their children nodes.
Although these functionalities are exploited differently across ontologies \cite{walk2017users},  
it is unclear how users navigate through the ontology structure. 
Thus, this section covers previous work on search and navigation on networks.

\para{Search}. \emph{Information Foraging}~\cite{pirolli1999information} assumes that people, when possible, modify their strategies or the structure of the environment to maximize their gain of valuable information. 
These patterns are also found in the way humans recall information from memory \cite{hills2012optimal}.
Similarly, \emph{berrypicking}~\cite{bates1989design}, a model of online searching, states that queries are not static, but rather evolve, and users commonly gather information in pieces instead of in one large set.

\para{Navigation}. 
\emph{\pagerank}~\cite{page1999pagerank} is the most popular method to measure the importance of web pages based on their incoming and outgoing links.
It relies on an imaginary surfer who is randomly \emph{clicking on links}, and eventually \emph{jumps} to any node in the network. The probability of following links is given by a \emph{damping factor}.
Multiple variations have been proposed for improving information retrieval systems, e.g., a biased PageRank \cite{haveliwala2003topic} to capture the importance of a page more accurately by taken topics into account or
a weighted \pagerank~\cite{xing2004weighted} to assign larger rank values to more popular pages (i.e., preferential attachment) instead of distributing the rank value of a page uniformly to all outgoing pages.
Geigl et al. suggest that the behavior of a \emph{random surfer} is almost similar to real users, as long as they do not use search engines \cite{geigl2015random}. They also find that classical navigation structures, such as navigation hierarchies or breadcrumbs, only exercise limited influence on navigation. 
Experiments in~\cite{singer2014detecting} reveal that memory-less \emph{\markovchain s} represent a quite practical model for human navigation on a page level. However, this assumption is violated when the analysis is expanded to a topical level.
Helic et al. identify certain configurations of \emph{decentralized search} that are capable of modeling human navigation in information networks~\cite{helic2013models}. Their findings suggest that navigation on such networks is a two phase process combined with the \emph{exploitation of the known} (i.e., goal-seeking) and the \emph{exploration of the unknown} (i.e., orientation).

\para{User Interfaces}.
Human navigation has also been 
studied for enhancing interfaces. For instance, \cite{furnas1986generalized}~explores \emph{fisheye} views to display large information structures such as programs and databases. The intuition behind this paradigm is that users often explore their neighborhood, and distant major landmarks in more detail. Similarly, Van Ham and Perer studied the \emph{search, show context, expand on demand} browsing model in \cite{van2009search}, and proposed techniques to design better graph visualization tools.

We propose \hoprank---a biased random walker---to model navigation on semantic networks.  
\hoprank~builds upon insights from information foraging \cite{pirolli1999information,hills2012optimal},  decentralized search \cite{helic2013models,kleinberg2002small} and \pagerank~\cite{page1999pagerank}.
More precisely, we replace the \emph{damping factor} by a \hopportation~vector to encode the probabilities of visiting each k-hop neighborhood.
The intuition here is that users browse semantically close terms more often than semantically distant ones.

\section{\bioportal}
\label{sec:bioportal}
There exist a large number of ontologies in the biomedical domain. They are highly specialized and therefore expensive to develop. To enable ontology adoption and reuse, effective support for browsing and exploring existing ontologies is required. Towards that goal, the National Center for Biomedical Ontology (NCBO) \cite{NCBO,musen2011national}
features BioPortal
\cite{BioPortal,noy2009bioportal,whetzel2011bioportal}---one of the leading repositories of biomedical ontologies on the Web---
containing currently more than $700$ ontologies with more than $9$ million ontology classes.
On \bioportal, practitioners and experts can access ontologies via Web services and Web browsers. The latter allows users to navigate ontologies by searching specific classes, or by directly browsing their concept hierarchies within a tree-like explorer \cite{walk2017users}.

\para{Ontologies. }
We propose to model human navigation on semantic networks using the structure of the underlying ontology. On \bioportal, ontologies are defined as directed networks, where nodes represent \emph{concepts} and edges \emph{isASubClassOf} relationships. Since such edges are usually non-cyclic and have a common root, these ontologies often form trees. 
\Cref{tbl:bioportal}~shows \nontologies~of the most visited ontologies in 2015\footnote{As ontologies can be edited over time, we work with their latest snapshots from \yearstudy.}. For instance, LOINC the largest ontology with \loincN~nodes, \loincE~edges, and \loincCC~connected components.

\begin{table}[b]
\centering
\caption{\textbf{Datasets}. This table illustrates network properties of \nontologies~of the most popular ontologies on \bioportal~in \yearstudy. Ontologies represent networks whose nodes refer to \emph{concepts} and edges \emph{isASubClassOf} relationships. Original number of nodes, edges, and connected components of ontologies are shown under N, E and cc, respectively. Properties of the largest connected component (LCC) of each ontology are shown under N', E', d' and T', where d' refers to the diameter and T' to the number of transitions.
}
\label{tbl:bioportal}
\scalebox{0.84}{
\begin{tabular}{@{}rl|rrr|rrrr@{}}
\toprule
\textbf{\#} & \textbf{Ontology} & \textbf{N} & \textbf{E} & \textbf{cc} & \textbf{N'} & \textbf{E'} & \textbf{d'} & \textbf{T'} \\ \midrule
1 & CPT & 13219 & 13235 & 3 & 13092 & 13110 & 15 & 44651 \\
2 & MEDDRA & 66506 & 31863 & 43493 & 22889 & 31738 & 8 & 42746 \\
3 & NDFRT & 35019 & 34504 & 522 & 32074 & 32080 & 24 & 22452 \\
4 & LOINC & 174513 & 152683 & 73518 & 100871 & 152558 & 13 & 6349 \\
5 & ICD9CM & 22534 & 22531 & 3 & 22407 & 22406 & 12 & 4434 \\
6 & WHO-ART & 1852 & 2997 & 3 & 1725 & 2872 & 4 & 2811 \\
7 & MESH & 165166 & 24182 & 145652 & 16947 & 21596 & 31 & 2623 \\
8 & ICD10 & 12446 & 11256 & 1190 & 11132 & 11131 & 10 & 2288 \\
9 & CHMO & 2966 & 3071 & 3 & 2964 & 3071 & 22 & 1423 \\
10 & HL7 & 10319 & 10600 & 1049 & 9146 & 10475 & 19 & 1374 \\
11 & OMIM & 81821 & 39359 & 44110 & 37587 & 39234 & 6 & 1291 \\ \bottomrule
\end{tabular}
}
\end{table}

\para{Transitions}.
We analyzed all HTTP requests made in \yearstudy~and extracted \nsessions~valid sessions (i.e., after filtering out sessions with less than \minsession~requests, and requests to ontologies or concepts which do not exist). Each session contains transitions (i.e., a sequence of visited concept pages) triggered by a single user (i.e., IP address) without breaks (i.e., pauses of at least \sessionbreak).
For simplicity, we only consider transitions within the largest connected component (LCC) of each ontology, and discard ontologies with less than \mintrans~transitions\footnote{Transitions within the LCCs of these ontologies represent \percentagetransitions~of all transitions.}. 
Overall, we found \nontologies~ontologies and \totaltransitions~
transitions between their concepts\footnote{We left out the popular SNOMEDCT ontology due to computational limitations.},
see \Cref{tbl:bioportal} for some key properties.

\begin{figure}[t]
    \centering
        \includegraphics[width=1.0\linewidth]{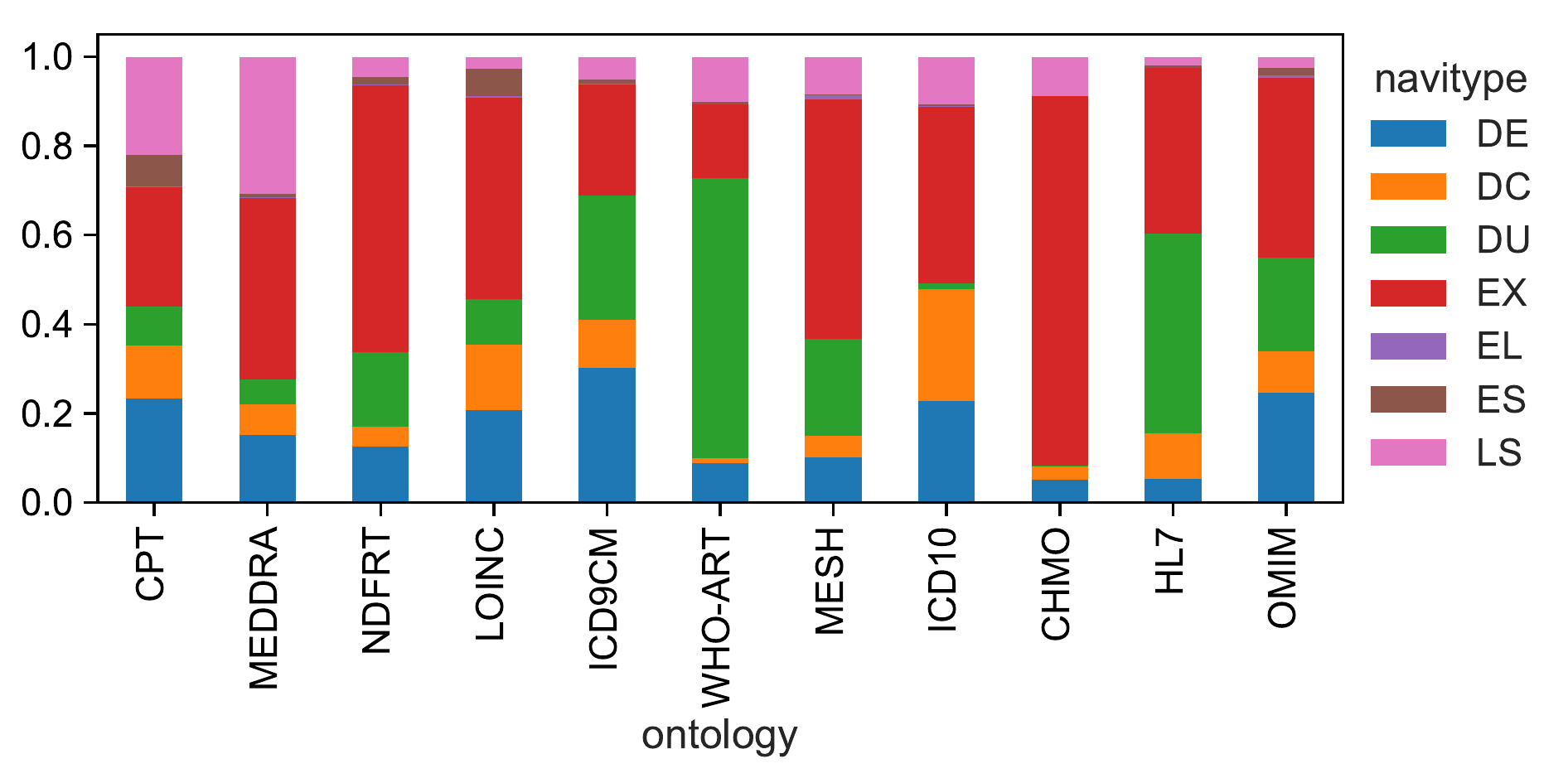}
    \setlength{\belowcaptionskip}{-5pt} 
    \caption{\textbf{Navigation Types}. Each bar shows the fraction of transitions within the LCC of each ontology. Stacked bars differentiate types of navigation: details (DE, blue), direct click (DC, orange), direct URL (DU, green), expand (EX, red), external link (EL, purple), external search (ES, brown) and local search (LS, pink). Most ontologies are mainly navigated by \emph{expanding} nodes within the tree-like explorer. }
    \label{fig:transitions}
\end{figure}

\para{Navigation Types}.
Based on the HTTP request headers, we inferred \nnavitypes~navigation types: Details (DE), Direct Click (DC), Direct URL (DU), Expand (EX), External Link (EL), External Search (ES), and Local Search (LS).
\textbf{DE}: are all clicks made within the \emph{Details} tab of a selected concept.
\textbf{DC}: are all clicks made on concepts within the tree-like explorer.
\textbf{DU}: refers to all concept requests without HTTP referrer (e.g., direct URL in the browser).
\textbf{EX}: considers all clicks on the (+) symbol of a concept, which triggers the expansion of the concept to show all its children nodes. Notice that this request is called only once, even if the symbol is clicked multiple times. The opposite behavior (collapse) is not considered\footnote{Collapse is a client-side functionality, and thus, it is not recorded in the log files.}.
\textbf{EL}: captures all requests coming from external websites that are not search engines. \textbf{ES}: are all requests coming from the top $10$ most popular external search engines such as Google and Yahoo. 
\textbf{LS}: are all requests made via the local search functionality of each ontology. Notice that this search is a 3-step process. First users type a keyword, then the system shows auto-suggestions and finally users click on one of the concepts shown in the auto-suggestion list. We only consider the final step a local search transition. 
\textbf{ALL}: includes all the above-mentioned types.
\Cref{fig:transitions} shows the distribution of transitions across navigation types for each ontology. In general, most traffic comes from expanding a concept (EX, $44\%$), followed by local search (LS, $17\%$), direct URL (DU, $16\%$) and details (DE, $14\%$). Surprisingly, direct clicks on concepts (DC) only represent $7\%$ of all transitions. 
This suggests that users spend substantial time expanding concepts before they find a concept of interest.

\section{\hoprank: A biased random walker}

\hoprank~models human navigation on semantic networks.
Imagine a random walker whose decisions on where to go next are biased towards specific \khop~neighborhoods. This bias is what we call \emph{\hopportation}, which encodes the probabilities of transitioning to each \khop~neighborhood.
In our model, navigation on networks can be explained as a 2-step process. First, a $k$-hop neighborhood of the current node $i$ is drawn from a categorical distribution. 
Second, a node $j$ is randomly chosen within that $k$-hop neighborhood. Note that this process holds only if the walker is fully or partially aware of the structure of the network (i.e., knows or can see it). Without this prerequisite, and if links are not preferred, then random jumps to random pages will be more plausible.
In comparison to the classic random walker with teleportation (e.g., \pagerank~\cite{page1999pagerank}), where its movements are constrained by the damping factor $\alpha$ (i.e., probability of following links), 
\hoprank~is constrained by a vector $\vec{\beta}$ containing $k$ different factors, which define the probabilities of going to each \khop~neighborhood from the current location.

\begin{figure}[t]
    \centering
    \subfigure[\% of dyads traversed per ontology]{
        \label{fig:visitedhops:dyads}
        \includegraphics[width=0.9\linewidth]{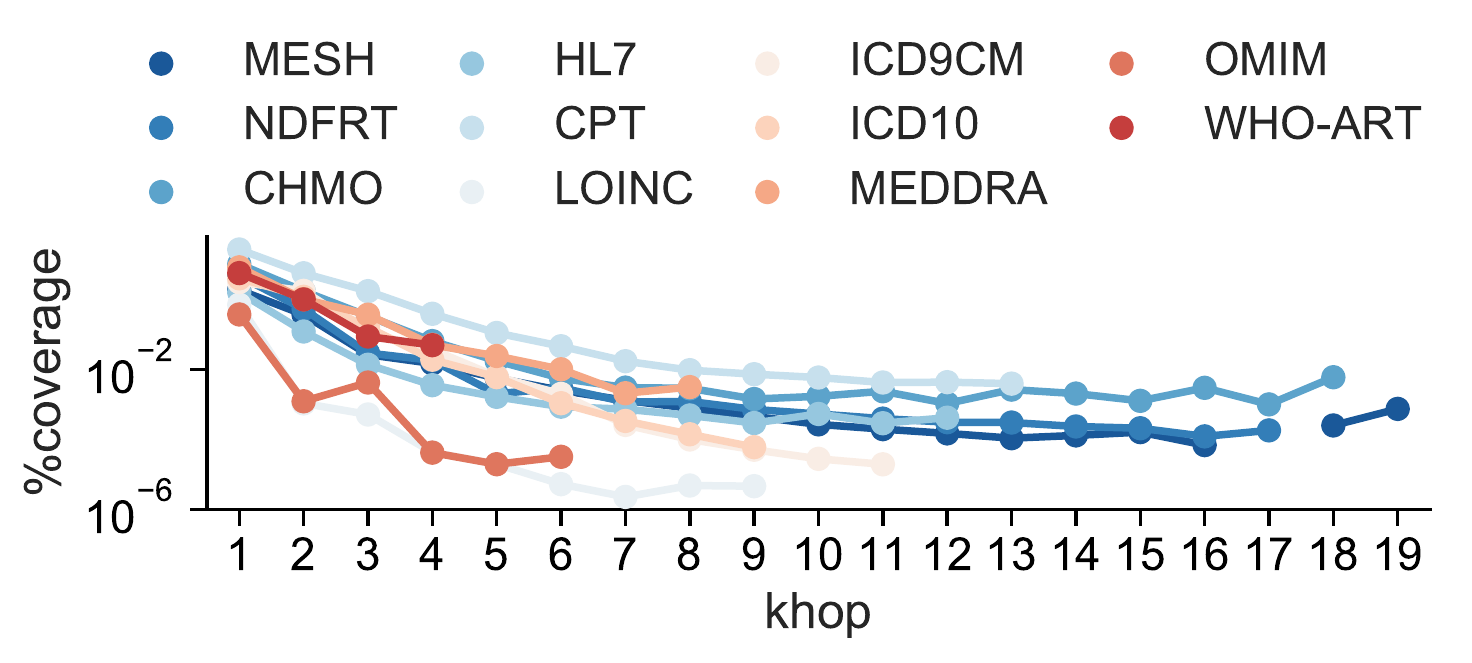}}
    \subfigure[Mean \% of transitions per navigation type]{
        \label{fig:visitedhops:transitions}
        \includegraphics[width=0.9\linewidth]{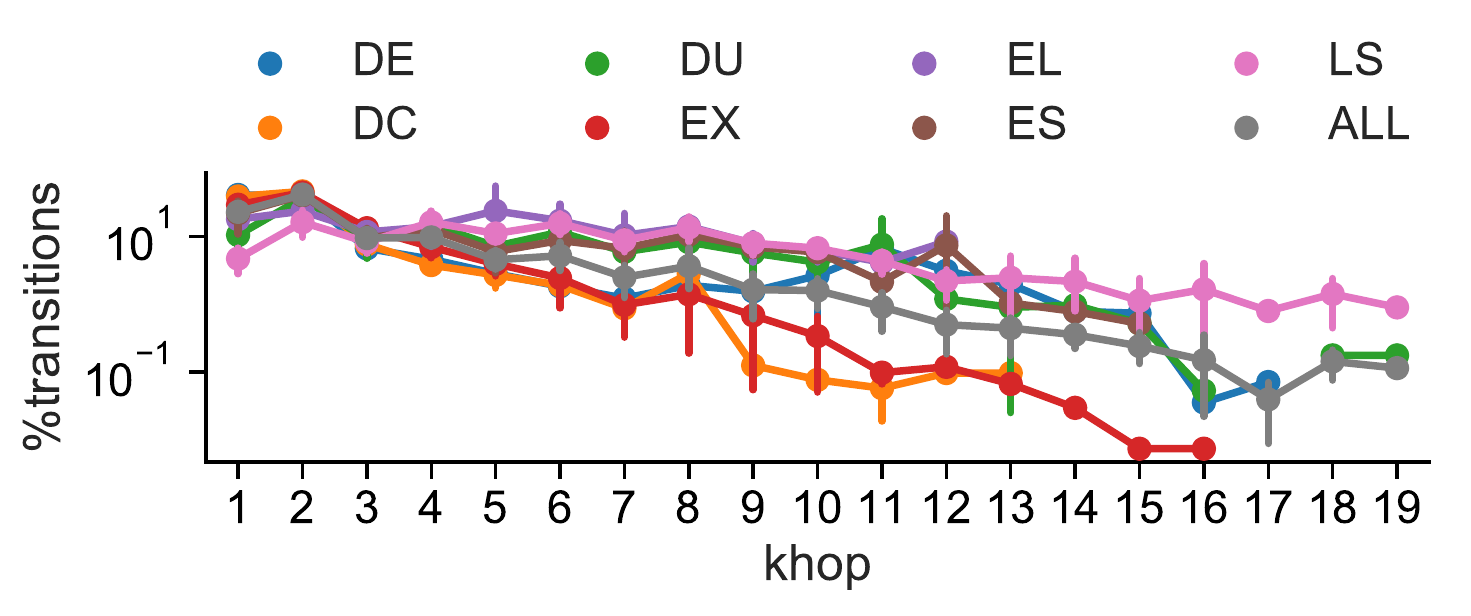}}
    \setlength{\belowcaptionskip}{-5pt}
    \caption{\textbf{Popularity of \khop s}.
    (a) Shows the percentage of dyads that are traversed per \khop~neighborhood. Lines represent ontologies and are sorted by their LCC diameter: In descendent order from MESH (dark blue) to WHO-ART (dark red). (b) Shows the distribution of transitions across \khop~neighborhoods per navigation type. Percentages are averages across ontologies, and error bars the respective standard deviation. While several k-hop distances are being traversed non-uniformly, most transitions happen across nearby nodes, especially when browsing (DE, DC, EX, ES) 2-hop neighbors. In contrast, non-browsing types (EL, LS, DU) tend to reach more distant nodes more frequently.}
    \label{fig:visitedhops}
\end{figure}

\para{Visited \khop~Neighborhoods on \bioportal}.
We aggregate ALL transitions by the shortest distance between two sequentially visited nodes.
This distance is referred to as \khop~neighborhood.
In \Cref{fig:visitedhops:dyads} we see that target nodes at large distances are less likely to be visited next.
This is expected, since---to some extent---larger distances enclose more branches, therefore more target candidates.
Note that ontologies are sorted by diameter in descendant order from MESH to WHO-ART.
Interestingly, users tend to hop as far as the ontology's diameter, for $d'\leq12$.
For instance, OMIM's diameter is $6$ (see \Cref{tbl:bioportal}), and $6$ is the maximum hop done by users. 
Otherwise, users (roughly) hop up to two-thirds of the ontology's diameter, for $d'>12$.
For example, MESH's diameter is $31$, and the largest hop reached is $19$.

\para{Transitions per \khop~Neighborhood on \bioportal}.
\Cref{fig:visitedhops:transitions} shows the average percentage of transitions across \khop~neighborhoods per navigation type. We see that users on average (ALL, grey) prefer to navigate through $2$-hop (41\%) and $1$-hop (23\%) neighbors. 
In particular, when navigation is triggered by direct clicks (DC, orange) and expand (EX, red). Notice their fast decay when $khop>8$.
Other types of navigation such as external link (EL, purple), and direct URL (DU, green)---which do not leverage the tree-like explorer---tend to reach concepts at larger distances more frequently. Notice their peaks at $khop=\{5,11\}$, respectively. 
Interestingly, when users opt for external search (ES, brown), they often click on 2-hop concepts, but also on 12-hop and 15-hop neighbors.
Intuitively, the details tab (DE, blue) helps users to click on nearby concepts at $khop \leq 2$, more often than local search (LS, pink), which is more likely to reach concepts at $khop \geq 2$.

\section{Models of Human Transitions}
\label{sec:modeling}
In this section, we formally introduce our \hoprank~model, and recap popular navigation models for comparison. 
We denote the transition probabilities, and \# of parameters  according to
{\hoprank}~and \nomodels~other models that we will use later on for model selection.

We formally represent an ontology\footnote{We focus on its largest connected component (LCC)} as a graph $G=(V,E)$, with $V=(v_1,\dots v_n)$ being a set of $N$ nodes, and $E=\{(v_i,v_j)\}\in {V\times V}$ a set of undirected edges\footnote{Directionality of edges is omitted to calculate shortest paths between all pair of nodes.}. The ontology structure is captured by the adjacency matrix $A_{N\times N} = {a_{ij}}$, where $a_{ij}$ is $1$ if the link exists, $0$ otherwise. 
Transitions are represented by the transition matrix $T_{N\times N} = {t_{ij}}$, where $t_{ij}$ represents the number of transitions between source node $i$ and target node $j$.

\para{\hoprank}. 
Given the \ontoportation~vector $\vec{\beta}$, the probability of reaching a \khop~neighborhood is denoted by factor $\beta_k \in \vec{\beta}$. 
$M_k$, the stochastic $k$-hop matrix, describes all nodes $j$ with a shortest distance $k$ from $i$. \hoprank~uniformly distributes $\beta_k$ across all nodes $j$ at distance $k$. The limits of \khop~neighborhoods go from $1$ (direct edges), to $d'$, the diameter of the ontology $G$. Noise $\beta_0 = 1-\sum_{k=1}^{d'} \beta_{k}$ is added to allow for random jumps and self-loops.  \Cref{fig:toy-example:hopportation} illustrates how the \ontoportation~vector is computed from the transition counts. \emph{Number of model parameters: $d'+1$}.

\begin{equation}
    \label{eq:hoprank}
    P_{HR} = \beta_1 \pmb M_1 + \beta_2 \pmb M_2 + \dots + \beta_k \pmb M_k + \frac{\beta_0}{N}
\end{equation}

\para{Preferential Attachment~(PA)}. 
Given the degree matrix $D_{N\times N} = {d_{ij}} = {d_j}$, where ${d_j}$ represents the degree of the target node $j$.
The probability of moving from $i$ to $j$ is proportional to the degree of $j$. 
\emph{Number of model parameters: $0$}.

\begin{equation}
    \label{eq:preferential}
    P_{PA} = \pmb D
\end{equation}

\para{\gravi~(Gr)}. 
Given the matrix $S_{N\times N} = {(sp(i,j)+\epsilon)^2}$, where $sp(i,j)$ denotes the shortest path between nodes $i$ and $j$. 
The probability of navigating from $i$ to $j$ is proportional to the degree of node $j$ and inversely proportional to the square distance between $i$ and $j$. We add a smoothing factor $\epsilon$ to avoid overflows when dyads are disconnected. In such cases, we set $\epsilon$ to the diameter $d'$ of $G$ plus $1$, to consider these jumps with a very low probability. Similarly, we set the diagonal (i.e., self-loops) to $\epsilon=d'+2$.
\emph{Number of model parameters: $0$}.

\begin{equation}
    \label{eq:gravitational}
    P_{Gr} = \frac{\pmb D}{ S }
\end{equation}

\para{Random Walker~(RW)}.
Given the damping factor $\alpha$ (i.e., probability of following links), the probability of visiting a node $j$ is proportional to $\alpha$ divided by the degree of the source node $i$, plus a random choice equally distributed among all nodes. Depending on the $\alpha$ value, a random walker can model four different behaviors: \textbf{(i)} $\alpha=0.0$: random jumps only, \textbf{(ii)} $\alpha \approx 1.0$: navigation over links only, \textbf{(iii)} $\alpha=0.85$: \pagerank~using the commonly used damping factor for navigating the Web \cite{brin1998anatomy}, and \textbf{(iv)} the empirical \pagerank~which learns the parameter $\alpha$ from the transitions data.
\emph{Number of model parameters: $1$ if empirical, $0$ otherwise.}

\begin{equation}
    \label{eq:pagerank}
    P_{PR} = \alpha \pmb A + \frac{(1-\alpha)}{N}
\end{equation}

\para{Markov Chain~(MC)}.
We assume that moving to the next node follows a Markov process. Therefore, the probability of moving to a node $j$ only depends on the current node $i$. These probabilities represent the maximum likelihood, learned from the transition matrix $T$. Thus, the probability of visiting node $j$ from node $i$ is proportional to the number of transitions $t_{ij}$.
\emph{Number of model parameters: $N \times (N-2)$}.

\begin{equation}
    \label{eq:markov}
    P_{MC} = \pmb T
\end{equation}

Note that $\pmb M$, $\pmb A$, and all $P_{*}$ from \Cref{eq:hoprank,eq:preferential,eq:gravitational,eq:pagerank,eq:markov} are right stochastic matrices (i.e., each row must sum to $1$).

\section{Experiments}
\label{sec:experiments}
In this section, we compare the performance of \hoprank~to the baselines on synthetic and real-world networks. 

\subsection{Model Selection}

For comparing the models, we employ the Bayesian Information Criterion (BIC)~\cite{schwarz1978estimating} to select the best, i.e., lowest BIC score. BIC evaluates \emph{log-likelihoods} $LL$ (i.e., how likely our transitions are for a given model) and takes into account the \emph{number of model parameters} and \emph{observations} (i.e., \# of transitions) to avoid over-fitting.
\begin{align}
    \label{eq:bic}
    BIC &= -2 \cdot LL + nparams \cdot log(nobservations),\\
    LL &= \sum_{i=1}^{N} \sum_{j=1}^{N} t_{ij} \cdot log(p_{ij}),
\end{align}

where $t_{ij}$ represents the actual number of transitions from node $i$ to node $j$, and $p_{ij}$ the probability of transitioning from node $i$ to node $j$ for a given model.

\subsection{Synthetic Network}
\para{Setup}. The underlying network (structure) is a binary tree composed by $N=7$ nodes and $|E|=6$ edges as shown in \Cref{fig:toy-example:sequence}. Transitions (curved-thick edges) are biased towards 2-hop and 4-hop neighborhoods.
These biases are reflected in the \hopportation~vector shown in \Cref{fig:toy-example:hopportation}. 

\begin{figure}[t]
    \centering
    \includegraphics[width=0.40\textwidth]{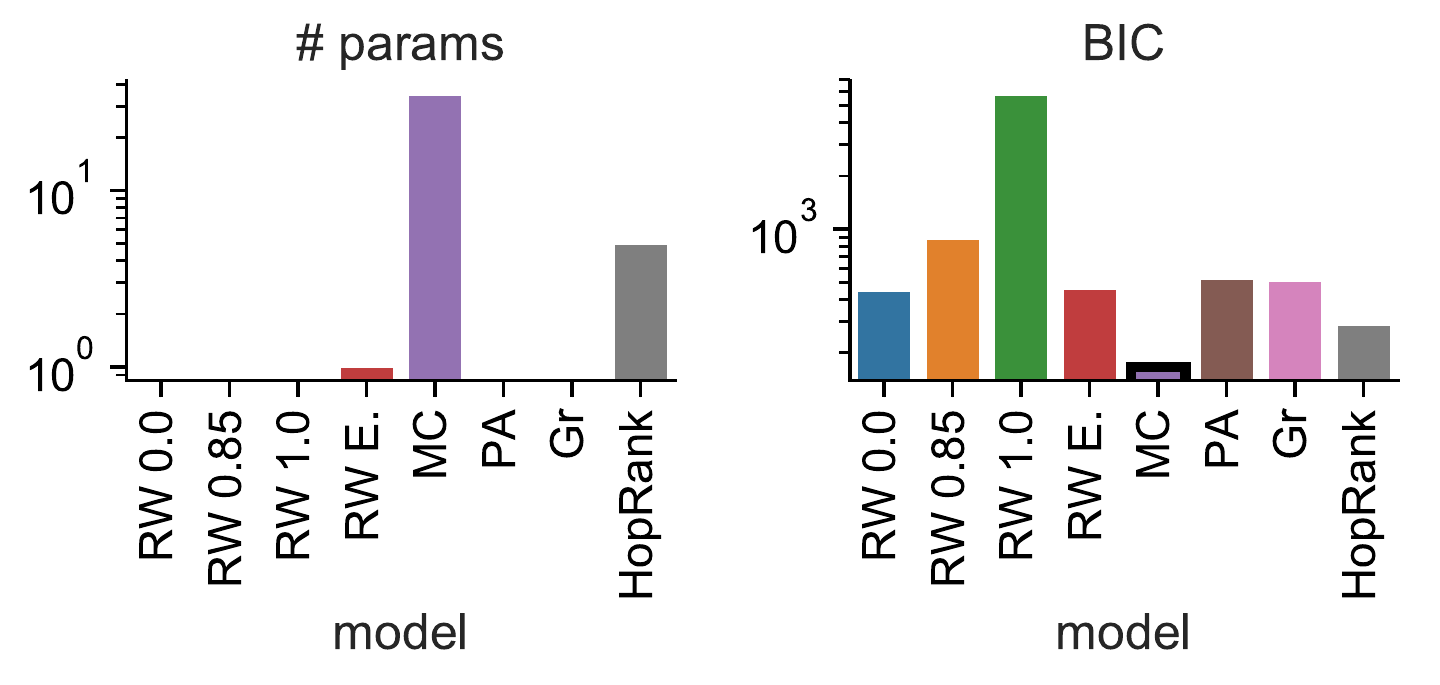}
    \setlength{\belowcaptionskip}{-5pt} 
    \caption{\textbf{Results on Synthetic Network from \Cref{fig:toy-example}}. 
    X-axis maps the models at interest. (a) Number of parameters inferred by each model. (b) BIC: 
    The lower the score, the better the model explaining the data. In this example, navigation is best described by \markovchain~followed by \hoprank.}
    \label{fig:toy-example:results}
\end{figure}

\para{Results}. Probabilities inferred using \Cref{eq:hoprank} are depicted in \Cref{fig:toy-example:probabilities}. \Cref{fig:toy-example:results} (left) shows the number of parameters inferred by each model. While the \markovchain~model (MC) requires 35 parameters, \hoprank~only needs $5$. The empirical \pagerank~(RW E.) learned a damping factor of ${\alpha=0.01}$. This means that users are $1\%$ likely to follow links. In \Cref{fig:toy-example:results} (right) we see the comparison of models using BIC scores. In this synthetic network, transitions are best described by the \markovchain~model because model parameters (i.e., maximum likelihood) are proportional to the actual transition counts per dyad, and the data structure is very small\footnote{Therefore, number of parameters does not play a very important role in BIC.}. In spite of that, \hoprank~is the second best model and describes navigation better than random (RW 0.0).

\subsection{Medical Dictionary for Regulatory Activities Terminology (MEDDRA)}
\para{Setup}.
MEDDRA\cite{MEDDRA}
is one of the the largest ontologies in our dataset (see \Cref{tbl:bioportal}). After pre-processing, its largest connected component (LCC) consists of \meddranodes~nodes and \meddratrans~transitions.

\begin{figure*}[t]
    \centering
    \subfigure[HopPortation vectors]{
        \label{fig:meddra:beta}
        \raisebox{15mm}{
        \includegraphics[width=0.28\textwidth]{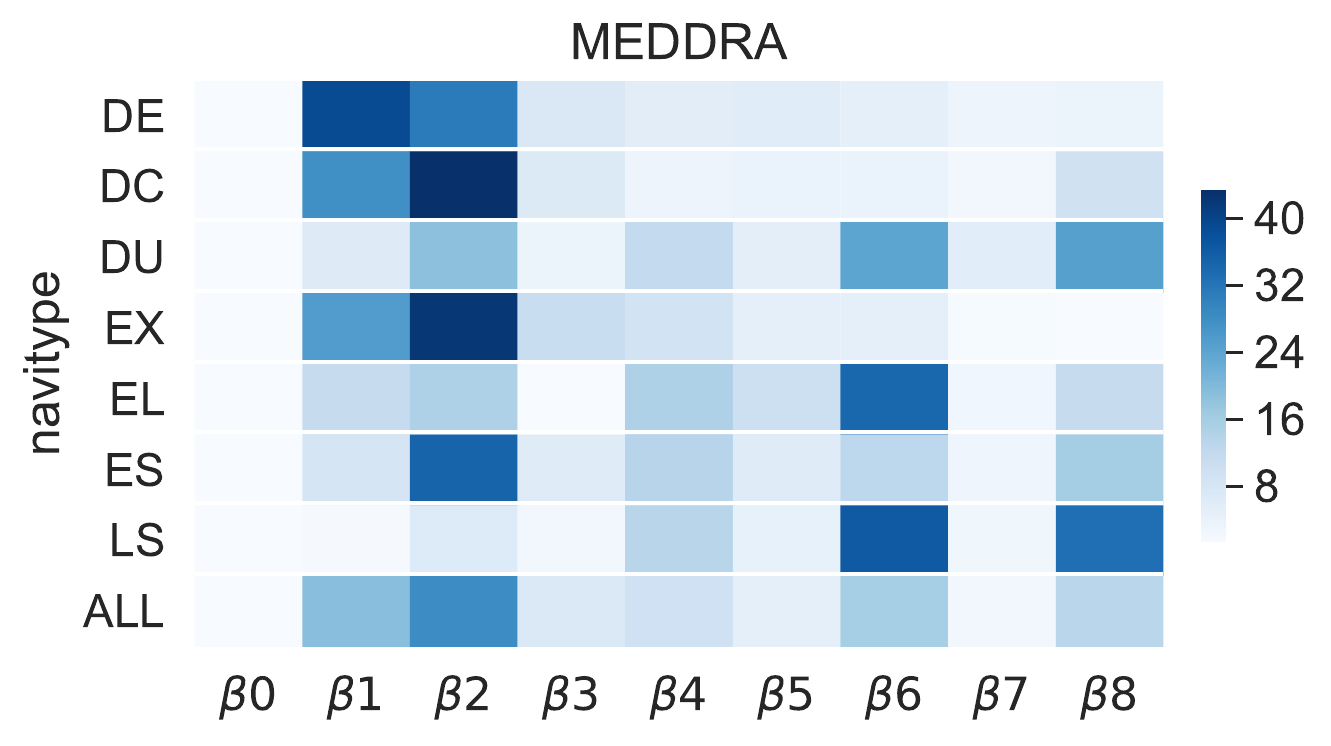}}}
    \subfigure[Model selection]{
        \label{fig:meddra:bic}
        \includegraphics[width=0.65\textwidth]{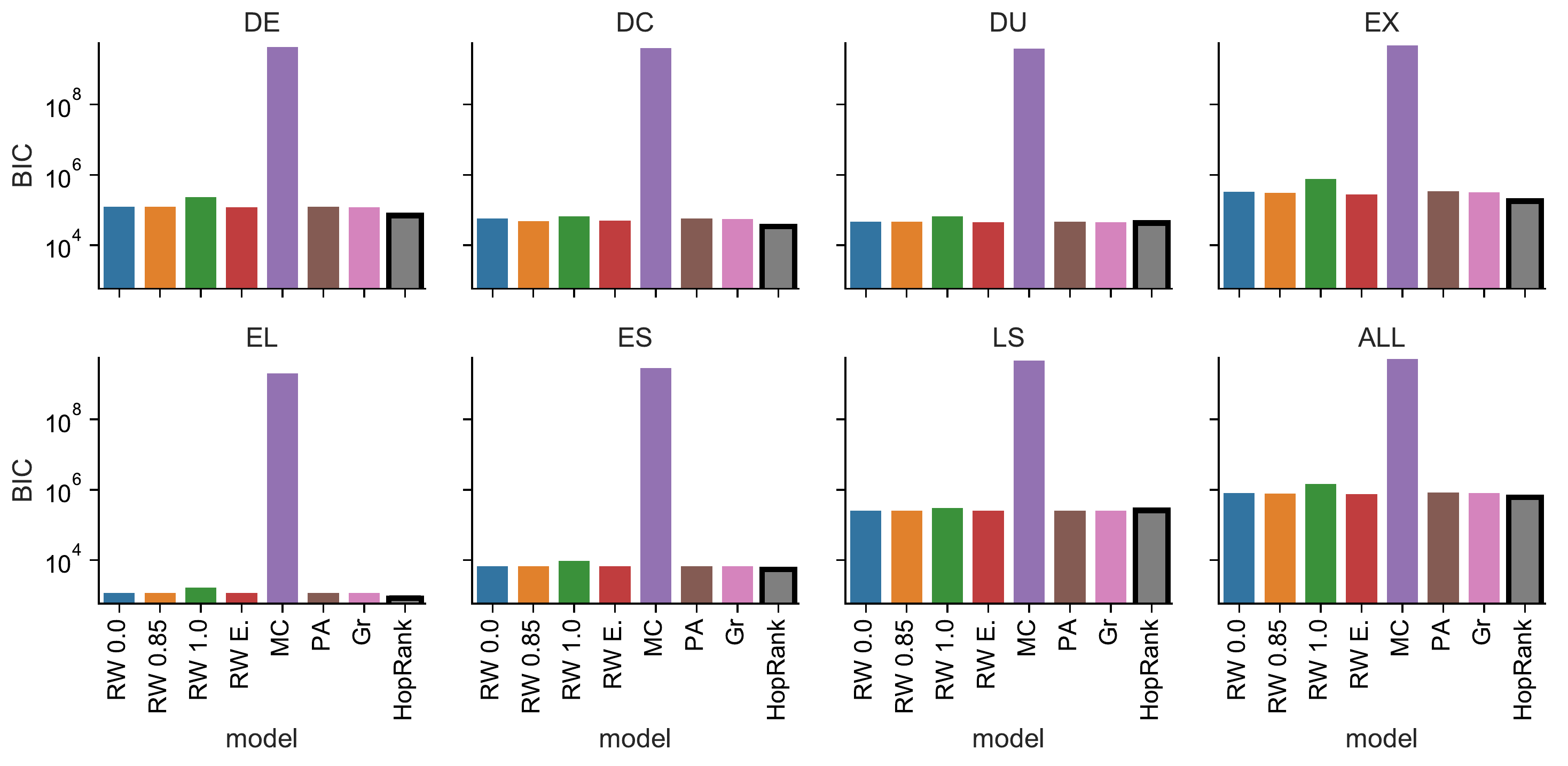}}
    
    \setlength{\belowcaptionskip}{-8pt} 
    \caption{\textbf{Results on MEDDRA}. (a) This heatmap shows the \hopportation~vectors learned from the transitions in MEDDRA. Cells represent the probabilities of visiting a certain \khop~neighborhood (column) by a given navigation type (row). In general, 2-hop and 1-hop neighborhoods are more likely to be visited next, regardless of navigation type (ALL). However, distant hops are preferred through \emph{direct URLs} (DU), \emph{external links} (EL), and \emph{local search} (LS). (b) This figure shows the comparison of models across navigation types using BIC scores. We see that \hoprank~outperforms all baseline models.}
    \label{fig:meddra}
\end{figure*}

\para{Results}. 
\Cref{fig:meddra:beta} shows the \hopportation~vectors learned for each type of navigation in MEDDRA. We see that users mainly navigate through 1, 2, 6, and 8-hop neighbors.
For instance, transitions through direct clicks---on a concept (DC), its details (DE) or expand (EX)---mainly follow 1-hop and 2-hop neighbors. However, when transitions are triggered by direct URLs (DU), local search (LS) or external links (EL), users tend to reach distant target nodes (i.e., 6-hop and 8-hop neighbors).
\Cref{fig:meddra:bic} shows the ranking of models according to BIC scores (lower is better). We see that in MEDDRA all types of navigation are best explained by \hoprank.

\subsection{Top\nontologies~Ontologies in \bioportal}
\para{Setup}. We fit \hoprank~and the baseline models to all transitions by ontology and navigation type. These represent \validtrans~transitions coming from the \nontologies~ontologies described in \Cref{tbl:bioportal}.

\begin{figure}[t]
  \centering
  \includegraphics[width=0.38\textwidth]{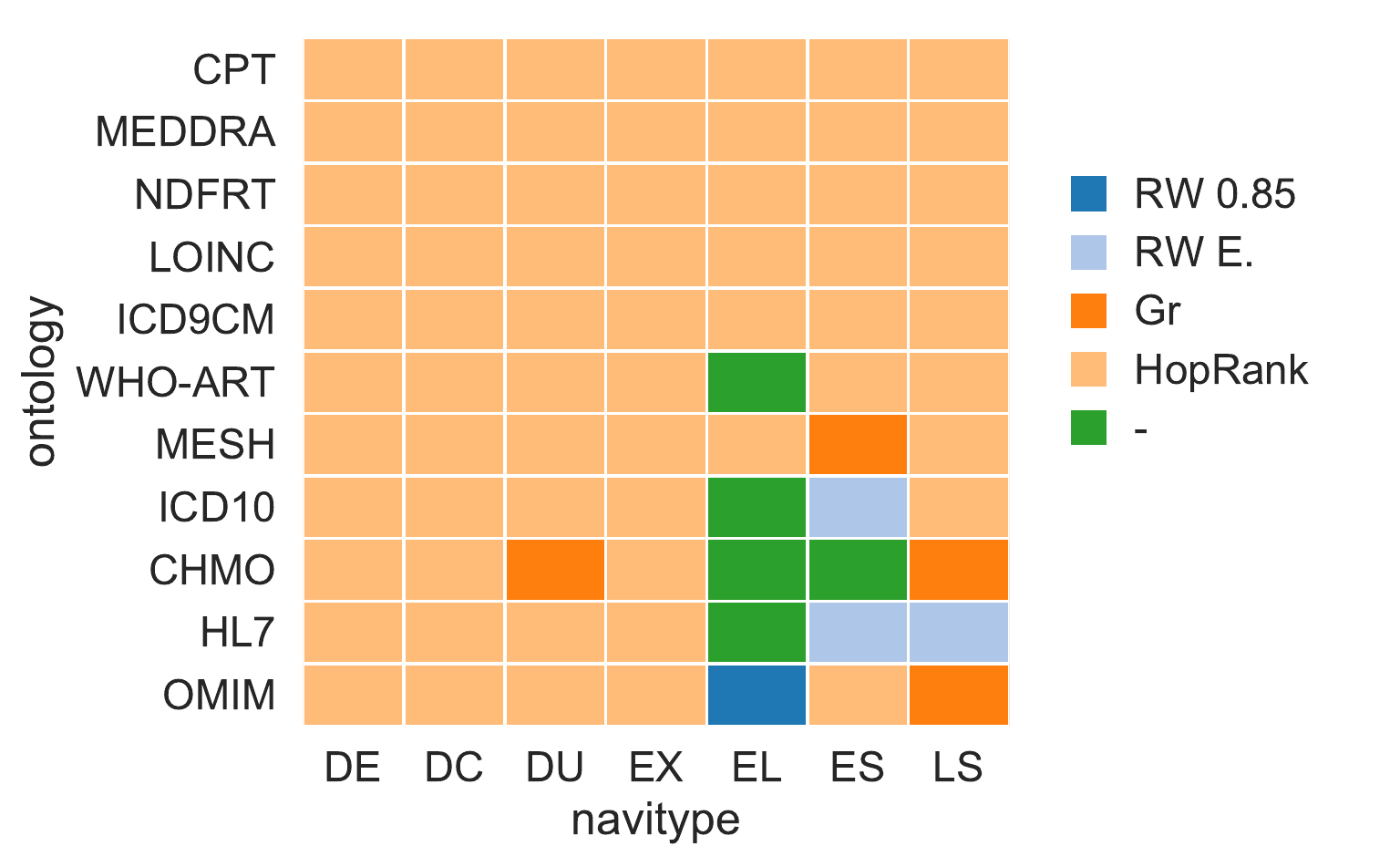}
  \setlength{\belowcaptionskip}{-8pt} 
  \caption{\textbf{Model Selection on \bioportal}. This heatmap highlights the model---with lowest BIC score---that best describes the \# of transitions per ontology and navigation type. \hoprank~outperforms the other models \hoprankoutperfoms~of the time, especially when browsing concepts via details (DE), direct click (DC) and expand (EX). When transitions are scarce (i.e., the other 11\%), BIC penalizes \hoprank~since it has more parameters than the other models (except \markovchain).
  }
  \label{fig:bioportal:modelselection}
\end{figure}

\para{Results}.
In \Cref{fig:bioportal:modelselection} we highlight the model that explains the number of transitions per ontology and navigation type best (i.e., the model with lowest BIC score). 
Ontologies are sorted by their number of transitions from CPT (largest) to OMIM (smallest). \hoprank~outperforms the other models \hoprankoutperfoms~of the time, especially when users browse directly---regardless of the ontology---the tree-like explorer via clicks (DC), details (DE) and expand (EX). When there are not enough observations (i.e., the number of transitions is small), the other models tend to outperform \hoprank~due to the fact that the other models require fewer parameters and/or it is less likely to find transitions across different \khop~neighborhoods. This is the case for $6$ ontologies in certain navigation types. For instance, we found $5$ external search (ES) transitions in MESH which are best described by the \gravi~model (Gr). Even though \hoprank~was a better candidate (i.e., higher log-likelihood), BIC penalized it for having more parameters ($nparams_{\hoprank}=32 > nparams_{Gr}=0$).
Notice that we model navigation in ontologies with at least $2$ transitions. Ontologies that do not fulfil this condition per navigation type are marked as green cells ``-''.

\section{Discussion and Future Work}
\label{sec:discussion}
In this section, we discuss decisions made for data processing, and future directions that can be pursued to improve our results.

\para{Largest Connected Component (LCC)}.
Surprisingly, ontologies on \bioportal~may have multiple connected components. In those cases, only the branch connected to the root \emph{owl:Thing} is shown at first in the tree-like explorer. Disconnected (and hidden) nodes or branches need to be accessed from external pages or local search.
For simplicity, we opted to work with the LCC of each ontology with the cost of removing 
$20\%$ of all transitions. Future work should consider the whole network to study the tradeoffs between number of transitions and random teleportation.  

\para{\hoprank~Extensions}.
More extensions based on network properties or similarity measures between nodes could improve our results. For instance, considering ontologies as directed graphs, and assuming that navigation is not only constrained by distance but also directionality: top-down or bottom-up.

\para{Other Types of Networks}.
Even though this paper targets semantic networks, 
we believe that \hoprank~can be utilized to model human navigation in other networks, such as the Web or cities. The only assumption required is that users must have background knowledge of the underlying network they are surfing/traveling in.

\section{Conclusions}
\label{sec:conclusions}
In this paper, we introduced the concept of \emph{\hopportation} which states that users---navigating a known or visible network---are biased towards certain \khop~neighborhoods. This is a variation of \pagerank, where we assume that teleportation is not fully random but rather distributed non-uniformly across different neighborhoods.
We proposed \emph{\hoprank}---a biased random walker---to model navigation on semantic networks. Our findings on \bioportal~
suggest that semantic structure (i.e., shortest path) influences navigation on networks. In particular, users tend to be biased towards certain \khop~neighborhoods depending on the type of navigation. For instance, when manually browsing the tree-like explorer, users tend to hop to nearby concepts, whereas far-away concepts are more likely to be reached by non-browsing types such as external links. These results advance our understanding of how ontologies are actually navigated and consumed, and help to develop and improve user interfaces for ontology exploration.

\para{Acknowledgements.} We would like to thank Tania Tudorache, John Graybeal, Matthew Horridge, Clement Jonquet, Maulik Kamdar, Alex Skrenchuk, Marcos Oliveira, Fabian Fl{\"o}ck, Reinhard Munz, Dimitar Dimitrov, Indira Sen, Mattia Samory and the three anonymous reviewers for their time and suggestions to improve the quality of the paper. This work was funded by DFG German Science Fund research projects ``KonSKOE'' and ``PoSTs II''.

\bibliographystyle{ACM-Reference-Format}
\balance

\end{document}